\documentclass[notitlepage,superscriptaddress,twocolumn,showpacs]{revtex4-1}
\usepackage{hyperref}
\usepackage{graphicx,subfigure}

\begin{document}
\title{Quantum Fisher information and coherence in  one-dimensional $XY$ spin models with
Dzyaloshinsky-Moriya interactions}
\author{Biao-Liang Ye}
\email{biaoliangye@gmail.com}
\affiliation{Quantum Information Research Center, School of Physics and
Electronic Information, Shangrao Normal
University, Shangrao 334001, China}
\affiliation{Jiangxi Province Key
Laboratory of Polymer Preparation
and Processing, Shangrao 334001, China}
\author{Bo Li}
\affiliation{School of Mathematics and
Computer Sciences, Shangrao Normal
University, Shangrao 334001, China}
\affiliation{Max-Planck-Institute for
Mathematics in the Sciences, Leipzig 04103, Germany}
\author{Zhi-Xi Wang}
\affiliation{School of Mathematical Sciences,
Capital Normal University, Beijing 100048,
China}
\author{Xianqing Li-Jost}
\affiliation{Max-Planck-Institute for
Mathematics in the Sciences, Leipzig 04103, Germany}
\affiliation{School of Mathematics and Statistics, 
Hainan Normal University, Haikou 571158, China}
\author{Shao-Ming Fei}
\email{feishm@cnu.edu.cn}
\affiliation{Max-Planck-Institute for
Mathematics in the Sciences, Leipzig 04103, Germany}
\affiliation{School of Mathematical Sciences,
Capital Normal University, Beijing 100048,
China}

\begin{abstract}
	We investigate quantum phase
	transitions in $XY$ spin models using Dzyaloshinsky-Moriya (DM)
	interactions. We identify the quantum critical points via 
	quantum Fisher information and quantum coherence,
	finding that higher DM couplings
	suppress quantum phase transitions. However, quantum coherence (characterized by the $l_1$-norm and relative entropy)
	decreases as the DM coupling increases. Herein,
	we present both
	analytical and numerical results.
\end{abstract}

\date{\today}
\maketitle

\section{Introduction}

Quantum entanglement plays important roles in
quantum physics and quantum information processing \cite{Horodecki2009,Ekert1991,Long2002,Deng2003,Zhang2017a,Zhu2017,Wu2017,Sheng2017,Xie2016},
such as in quantum key distribution \cite{Ekert1991,Long2002}, quantum secure direct communication
\cite{Deng2003,Zhang2017a,Zhu2017,Wu2017}, quantum machine learning \cite{Sheng2017}.
Quantum phase transition,
a key element of condensed matter physics,
is a type of quantum fluctuation that
occurs in spin-chain systems at zero temperature.
Understanding
the connections between quantum entanglement and
critical behavior near a quantum phase transition
is especially important
\cite{Osterloh2002,Osborne2002,Vidal2003}.

In recent decades, additional quantum measures
beyond entanglement
have been proposed \cite{Ollivier2001,Henderson2001,Oppenheim2002,Luo2008,Modi2010}.
Quantum phase transitions have been
intensively studied using correlation
measures \cite{Modi2012,Adesso2016}.
In particular, spin-chain systems, including the
$XY$, $XXZ$, and LMG models,
have been used to
investigate quantum discord \cite{Liu2011,Dillenschneider2008,Sarandy2009,Werlang2010,Campbell2013},
one-norm quantum discord \cite{Montealegre2013}, and quantum deficit \cite{Ye2017},
among others \cite{Karpat2014,Misra2015,Czekaj2015,Li2016,Malvezzi2016,Radhakrishnan2017}.
Recently,
quantum coherence has
attracted considerable interest as a resource.
The $l_1$ norm and
relative entropy of coherence are very intuitive
and coherence quantifiers are easy to compute
and fully monotone in all possible coherence
resource theories \cite{Baumgratz2014}.

Quantum Fisher information (QFI) \cite{Braunstein1994} is a central quantity
in quantum metrology \cite{Giovannetti2006}, where
a linear interferometer 
is used to estimate the unitary dynamics \cite{Czekaj2015}.
QFI also plays a significant role in quantum detection and estimation as it provides a bound on the quantum estimation accuracy. In addition, QFI signaling of quantum phase transitions has received much attention \cite{Liu2016a}.

The $XY$ model with Dzyaloshinsky-Moriya (DM) interactions has been extensively investigated. However, only a few studies have considered this model from the perspective of understanding the role
played by quantum correlations in phase transitions
\cite{Liu2011,Altintas2012,Radhakrishnan2017}.
In Ref. \cite{Liu2011},
quantum phase transitions of the above mentioned model have also been investigated in terms of quantum concurrence, quantum discord, and classical correlations. Herein, we study the quantum phase transition of the $XY$ model with DM interaction using QFI, in which QFI is an intrinsic and ubiquitous quantity that plays significant roles in quantum metrology, such as in parameter estimation. We also consider the $l_1$ norm and relative entropy of coherence.

The remaining paper is organized as follows.
In Sect. II, we review the basic
concepts behind QFI and quantum coherence measures.
In Sect. III, we introduce the $XY$ model with
DM interactions, presenting the
two-spin reduced density matrices and two-point
correlation functions.
In Sect. IV, we provide both analytical and
numerical results for the $XY$ model,
finding that higher DM coupling
parameters suppress quantum phase transition
behavior. Finally, in Sect. V, we present
our conclusions.

\section{Preliminaries}
First, let us briefly review QFI and quantum coherence.

\emph{Quantum Fisher information}.
In general phase estimation
scenarios, the evolution of a quantum state,
given by the density matrix $\varrho$, under a unitary transformation
can be described as
$\varrho_\theta=e^{-i A\theta}\varrho e^{i A\theta}$,
where $\theta$ is the phase shift and $A$ is
an operator. The estimation accuracy for $\theta$
is limited by the quantum
Cram\'{e}r-Rao inequality \cite{Helstrom76,Holevo82}:
\begin{equation}
	\Delta\hat{\theta}\ge\frac{1}{\sqrt{\nu \mathcal{F}(\varrho_\theta)}},
\end{equation}
where $\hat{\theta}$ denotes the unbiased
estimator for $\theta$, $\nu$ is the number of times the measurement is repeated,
and $\mathcal{F}(\varrho_\theta)$ is the so-called
QFI.

QFI is defined as
\begin{equation}\label{f}
	\mathcal{F}(\varrho, A)=2\sum_{m,n}\frac{(p_m-p_n)^2}{(p_m+p_n)}|\langle m|A|n\rangle|^2,
\end{equation}
where $p_m$ and
$|m\rangle$ are 
the eigenvalues and eigenvectors, respectively, of the
density matrix $\varrho$, which is used
as a probe state to estimate $\theta$.

\emph{Quantum coherence}.
Quantum coherence is an important resource in
quantum information tasks \cite{Streltsov2017}.
A rigorous theory has been proposed to
determine a good indicator for measuring quantum coherence \cite{Baumgratz2014}.
Herein, we focus on the $l_1$ norm and relative entropy
of coherence.

(i) $l_1$ norm.
The $l_1$ norm of coherence
is defined as the sum of the absolute values of all
off-diagonal elements in the density
matrix $\varrho$:
\begin{equation}
	C_{l_1}(\varrho)=\sum_{i\ne j}|\varrho_{ij}|.
\end{equation}

(ii) Relative entropy.
The relative entropy of coherence (REC) is defined as
\begin{equation}\label{re}
	C_{REC}(\varrho)=S(\varrho_{\rm{diag}})-S(\varrho).
\end{equation}
Here, $\varrho_{\rm{diag}}$ is the diagonal
part of $\varrho$, and the function $S(\sigma)=-\rm{Tr}
\sigma\log_2\sigma$ is the von Neumann entropy
of the density matrix $\sigma$. In the following discussion,
we will use both coherence measures as informational
tools to study quantum phase transitions in the
$XY$ model with DM interactions.

\section{The $XY$ model with DM interactions}
The Hamiltonian of the Heisenberg $XY$ chain with
DM interactions under periodic boundary conditions
in the presence of
an external field is described as follows:
\begin{eqnarray}\label{model}
	H=&&\sum\nolimits_{i=1}^{N}\{J[(1+\gamma)\sigma_i^x\sigma_{i+1}^x
	+(1-\gamma)\sigma_i^y\sigma_{i+1}^y\nonumber\\
	&&+D(\sigma_i^x\sigma_{i+1}^y-\sigma_i^y\sigma_{i+1}^x)]-\sigma_i^z\},
\end{eqnarray}
where $\sigma_i^\alpha (\alpha=x, y, z)$
are the Pauli operators at the $i$th
lattice site, $N$ is the total number
of spins, and $J$ is the
inverse of the external transverse magnetic
field strength. The system is antiferromagnetic
when $J>0$ and ferromagnetic when $J<0$.
The anisotropy $\gamma$ lies in the interval
$[-1, 1]$, with the values $\gamma=0$ and
$\pm1$ corresponding to the $XX$
and Ising models, respectively.
Finally the factor $D$ represents the strength of the
antisymmetric DM interaction along the $z$
direction.
In the following, we consider the case $N\rightarrow \infty$ at
zero temperature.

\begin{figure*}[htbp!]
\includegraphics[width=6.3in]{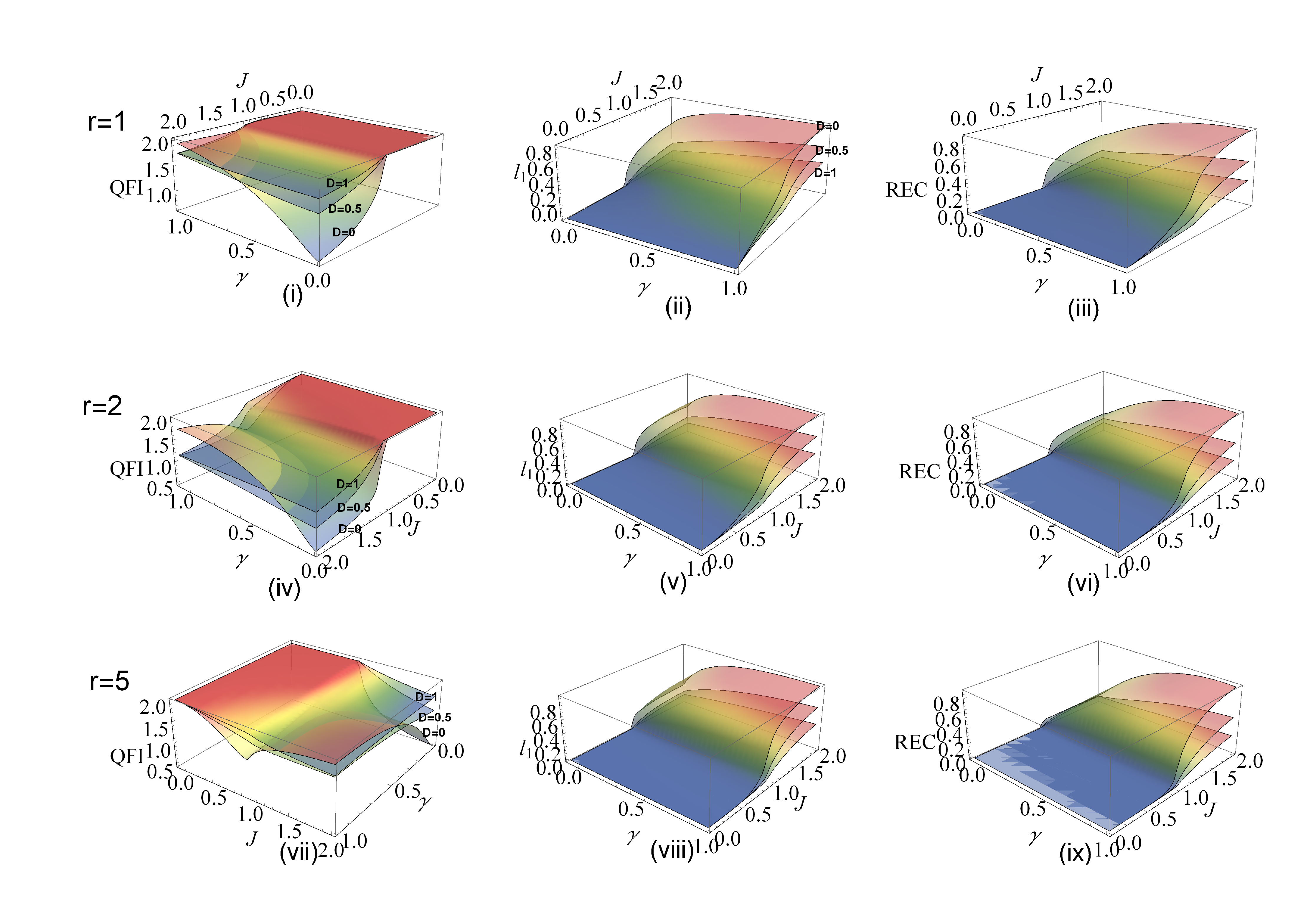}\\
\centering
\caption{(Color online) QFI, $l_1$ norm of coherence
and REC
for the
$XY$ model with DM interactions vs $\gamma$ and $J$,
for the DM parameter $D=0,0.5,1$.
The first (i,ii,iii), second (iv,v,vi),
and third (vii,viii,ix) rows correspond
to the nearest ($r=1$), next-nearest ($r=2$),
and fifth-nearest ($r=5$) neighbors,
respectively. The first (i,iv,vii), second
(ii,v,viii) and third 
(iii,vi,ix) columns correspond to the quantum Fisher information (QFI),
$l_1$ norm of coherence, and relative entropy of coherence (REC), respectively.
The DM interaction parameter $D$ is $D=0,0.5,1$ for
the top, middle and bottom surfaces, respectively,
in Fig. (ii), (iii), (v), (vi), (viii), and (ix).}
\label{fig1}
\end{figure*}

The $XY$ model has been solved exactly, and
the correlation functions were obtained from
\cite{Lieb1961,Perk1977}.
The $XY$ model with DM interactions can be solved
exactly using the Jordan-Wigner transformation \cite{Giamarchi2004}, as well \cite{Liu2011}.
The nonclassical correlation between two
spins at sites $i$ and $j$ can be
derived from their collective state, and
the model can be reduced to the two-spin density
matrix $\varrho(i,j)={\rm Tr}_{\backslash ij}\varrho$,
by tracing over all spins except those at
sites $i$ and $j$. The Hamiltonian's translational invariance ensures that the density matrix
$\varrho(i,j)=\varrho(i,i+r)$, where $r$ is
the distance between the spins.
This two-spin reduced density matrix exclusively depends
on $r$ and is independent
of the spins' actual locations. Thus, owing to the $U(1)$ invariance and symmetries
of the Hamiltonian Eq. (\ref{model}), 
the model's two-spin density matrix can be written as
\begin{eqnarray}\label{x}
\varrho(i,i+r)=\left(
\begin{array}{cccc}
	a_+&0&0&c_-\cr
	0&b&c_+&0\cr
	0&c_+&b&0\cr
	c_-&0&0&a_-
\end{array}
\right). 	
\end{eqnarray}
The entries in this density matrix
are given by two-point correlation functions,
\begin{eqnarray}
a_\pm&&=\frac14\pm\frac{\langle\sigma_i^z\rangle}2+\frac{\langle\sigma_i^z\sigma_{i+r}^z\rangle}4,\nonumber\\
	b&&=\frac{1-\langle\sigma_i^z\sigma_{i+r}^z\rangle}4,\nonumber\\
	c_\pm&&=\frac{\langle\sigma_i^x\sigma_{i+r}^x\rangle\pm\langle\sigma_i^y\sigma_{i+r}^y\rangle}4.
	\end{eqnarray}
The model's magnetization in the presence of an external
field is given by
\begin{equation}
	\langle\sigma_i^z\rangle=-\frac1\pi
	\int_0^\pi d\phi\frac{J(\cos\phi-2D\sin\phi)-1}{\Delta},
\end{equation}
where
\begin{equation}
\Delta=\sqrt{[J(\cos\phi-2D\sin\phi)-1]^2+J^2\gamma^2\sin^2\phi}.
\end{equation}

\begin{figure*}[htbp!]
\includegraphics[width=6.3in]{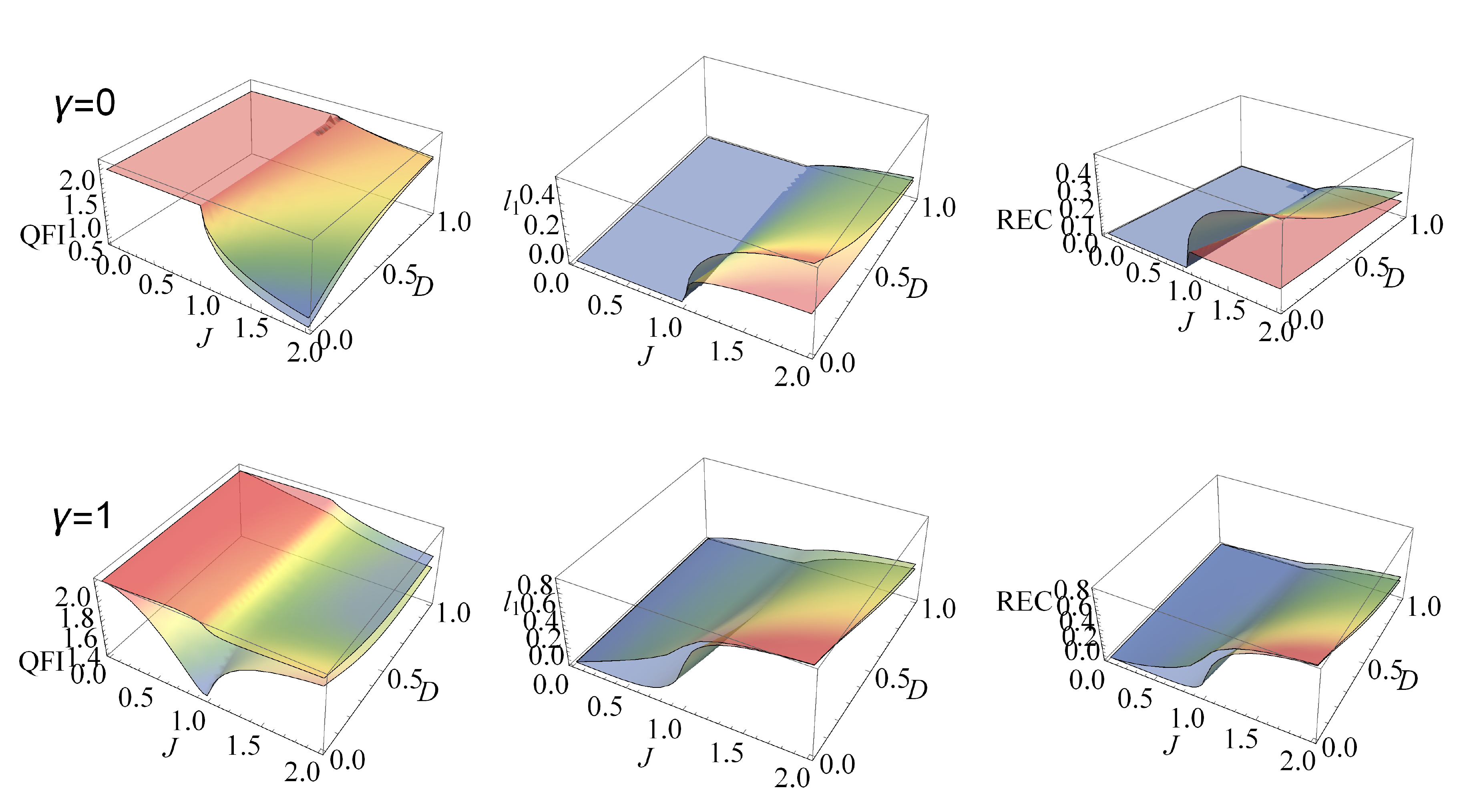}\\
\centering
\caption{(Color online) QFI, $l_1$ norm of coherence, and REC vs $J$ and $D$ for the $XX$ ($\gamma=0$) (first
row) and Ising ($\gamma=1$) (second
row) models. In all subfigures, the upper and lower surfaces correspond to the nearest ($r=1$) 
and fifth-nearest ($r=5$) neighbor cases, respectively.}
\label{xxising}
\end{figure*}

The two-point spin-spin correlation functions
corresponding to the $x$ and $y$ directions can be
computed from the determinants of the Toeplitz
matrices, and the one corresponding to the $z$ direction can be computed
from the magnetization, yielding
\begin{eqnarray}
	\langle\sigma_i^x\sigma_{i+r}^x\rangle=
	\left|
	\begin{array}{cccc}
	Q_{-1}&Q_{-2}&\dots&Q_{-r}\cr
	Q_0&Q_{-1}&\dots&Q_{-r+1}\cr
	\vdots & \vdots &\ddots&\vdots\cr
	Q_{r-2}&Q_{r-3}&\dots&Q_{-1}
	\end{array}
	\right|,
\end{eqnarray}
\begin{eqnarray}
	\langle\sigma_i^y\sigma_{i+r}^y\rangle=
	\left|
	\begin{array}{cccc}
	Q_{1}&Q_{0}&\dots&Q_{-r+2}\cr
	Q_2&Q_{1}&\dots&Q_{-r+3}\cr
	\vdots & \vdots &\ddots&\vdots\cr
	Q_{r}&Q_{r-1}&\dots&Q_{1}
	\end{array}
	\right|,
\end{eqnarray}
and
\begin{equation}
	\langle\sigma_i^z\sigma_{i+r}^z\rangle=\frac{\langle\sigma^z\rangle^2-Q_rQ_{-r}}4,
\end{equation}
where
\begin{eqnarray}
	Q_r=&&-\frac1\pi\int_0^\pi d\phi\frac{2\cos(\phi\cdot r)}{\Delta}[J(\cos\phi-2D\sin\phi)-1]\nonumber\\
	&&+\frac{\gamma}{\pi}\int_0^\pi d\phi\frac{2J\sin(\phi\cdot r)}{\Delta}\sin\phi.
\end{eqnarray}
Therefore, the model's QFI and quantum coherence 
can be
computed from the two-spin reduced density matrix Eq.(\ref{x}).

\section{Quantum Fisher information and quantum coherence
for the $XY$ model with DM interactions}

\subsection{Quantum Fisher information and quantum coherence}
Now, we study QFI for the $XY$ model with DM interactions.
Suppose
$\varrho$ is an arbitrary bipartite state,
whereas $\{A_\mu\}$ and $\{B_\mu\}$
are arbitrary complete sets of local
orthonormal observables of the two subsystems with respect to $\varrho$.
The QFI can then be written as \cite{li2013}
\begin{equation}\label{cal}
	\mathcal{F}=\sum_\mu \mathcal{F}(\varrho, A_\mu\otimes I
	+I\otimes B_\mu),
\end{equation}
which is also the global information
for $\varrho$.
It has been proved that
the value of QFI $\mathcal{F}$ given by Eq. (\ref{cal}) is independent of the choice
of local orthonormal bases \cite{li2013}, meaning
that it is an intrinsic quantity of
the composite system.
For a general two-spin system, the local
orthonormal observables
$\{A_\mu\}$ and $\{B_\mu\}$ can be defined as
\begin{equation}
	\{A_\mu\}=\{B_\mu\}
	=\frac{1}{\sqrt{2}}\{I,\sigma^x,
	\sigma^y,\sigma^z\}.
\end{equation}
Consequently, until $\varrho$ is given,
QFI $\mathcal{F}$ can be calculated from Eq. (\ref{f}).

For an $XY$ chain with DM interactions, after tedious computation
we can obtain the QFI for the two-spin state Eq. (\ref{x}):
\begin{widetext}
\begin{eqnarray}\label{ana}
\mathcal{F}=&&\frac{\left(\langle\sigma_i^x\sigma_{i+r}^x\rangle-\langle\sigma_i^y\sigma_{i+r}^y\rangle\right)^2}{1+\langle\sigma_i^z\sigma_{i+r}^z\rangle}+
	[\left(3 \langle\sigma_i^z\rangle^2+\langle\sigma_i^z\sigma_{i+r}^z\rangle^2-2 \langle\sigma_i^z\sigma_{i+r}^z\rangle\right)\left(\langle\sigma_i^x\sigma_{i+r}^x\rangle+\langle\sigma_i^y\sigma_{i+r}^y\rangle\right)\nonumber\\
	&&+\left(1-2 \langle\sigma_i^z\sigma_{i+r}^z\rangle\right) \left(\langle\sigma_i^x\sigma_{i+r}^x\rangle^2+\langle\sigma_i^y\sigma_{i+r}^y\rangle^2\right)+2 \left(\langle\sigma_i^z\rangle^2+\langle\sigma_i^z\sigma_{i+r}^z\rangle^2-2 \langle\sigma_i^z\rangle^2 \langle\sigma_i^z\sigma_{i+r}^z\rangle\right)\nonumber\\
	&&+\left(\langle\sigma_i^x\sigma_{i+r}^x\rangle^3+\langle\sigma_i^y\sigma_{i+r}^y\rangle^3\right)
	]/[\left
	(1+\langle\sigma_i^x\sigma_{i+r}^x\rangle\right)
	\left(1+\langle\sigma_i^y\sigma_{i+r}^y \rangle\right)
	-\langle\sigma_i^z\rangle^2].
\end{eqnarray}
\end{widetext}

Next, we compute the model's coherence.
As the two-spin reduced density matrix Eq. (\ref{x}) was computed in the $\sigma_z$ basis,
we study the coherence in the same basis.
Calculating the $l_1$ norm is straightforward:
\begin{equation}
	C_{l_1}=|\langle\sigma_i^x\sigma_{i+r}^x\rangle|.
\end{equation}
Using the relative entropy formula Eq. (\ref{re}), we have
\begin{equation}
	C_{REC}=\sum_{k=0}^1 (\xi_k\log\xi_k+\eta_k\log\eta_k-\zeta_k\log
	\zeta_k)-2\varepsilon\log\varepsilon,
\end{equation}
where
\begin{equation}
	\xi_k=[1-\langle\sigma_i^z\sigma_{i+r}^z\rangle
	+(-1)^k(\langle\sigma_i^x\sigma_{i+r}^x\rangle+\langle\sigma_i^y\sigma_{i+r}^y\rangle)]/4,
\end{equation}
\begin{eqnarray}
 	\eta_k=&&[
 	(-1)^k\sqrt{4\langle\sigma_i^z\rangle^2+(\langle\sigma_i^x\sigma_{i+r}^x\rangle-
 	\langle\sigma_i^y\sigma_{i+r}^y\rangle)^2}\nonumber\\
 	&&+1+\langle\sigma_i^z\sigma_{i+r}^z\rangle]/4,
\end{eqnarray}
\begin{eqnarray}
	\zeta_k=[1+\langle\sigma_i^z\sigma_{i+r}^z\rangle+(-1)^k2\langle\sigma_i^z\rangle]/4,
\end{eqnarray}
and
\begin{equation}
	\varepsilon=[1-\langle\sigma_i^z\sigma_{i+r}^z\rangle]/4.
\end{equation}

\begin{figure*}[htbp!]
\includegraphics[width=6.3in]{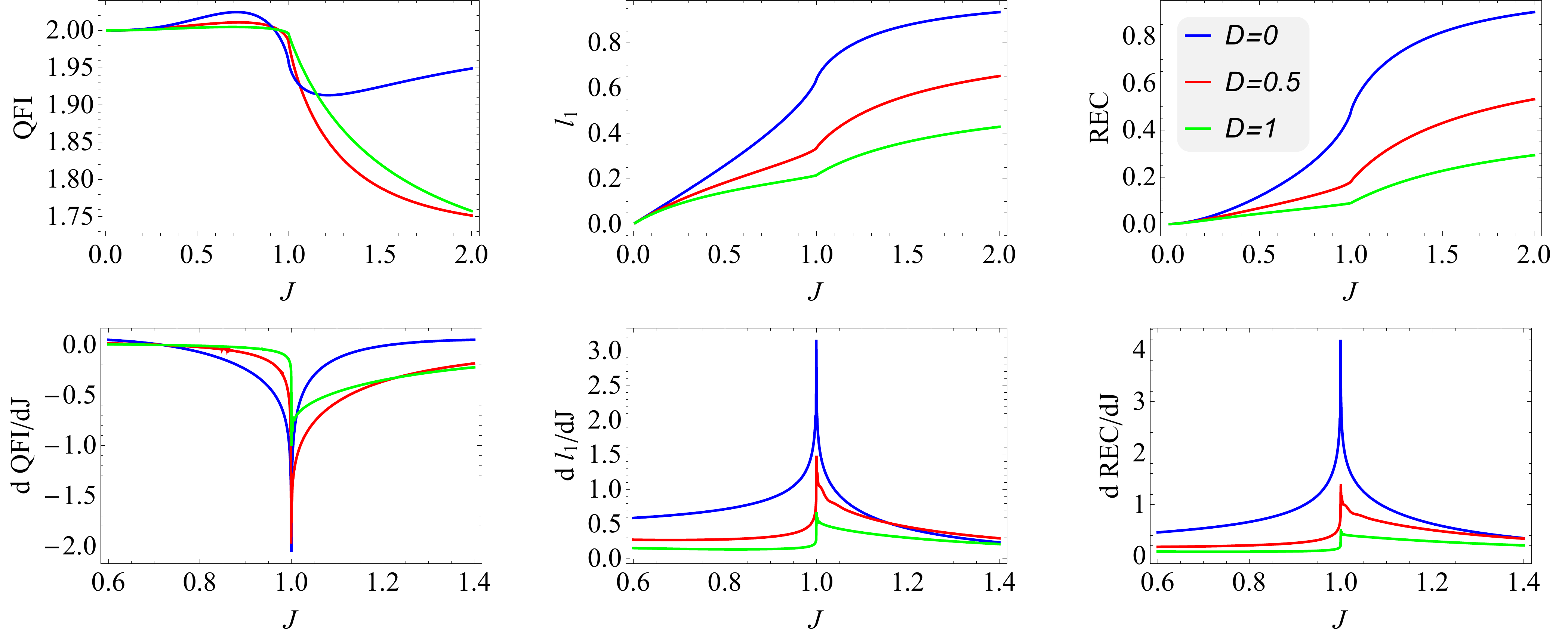}\\
\centering
\caption{(Color online) QFI, $l_1$ norm of coherence,
REC, and their derivatives vs $J$
for the nearest neighbor ($r=1$) spins in
the $\gamma=1$
$XY$ model with DM interactions.
The blue, red, and green lines correspond to $D=0,0.5,1$, respectively.}
\label{fig2}
\end{figure*}

\subsection{Quantum phase transitions}
Next, we show how quantum phase transitions can be detected in the $XY$ model with DM interactions.
Fig.~\ref{fig1} plots the QFI, $l_1$ norm of coherence, and
REC for $XY$ chains with different DM interaction values $D$, namely $0$, $0.5$, and $1$.
These show that the QFI in the 
$J<1$ region differs from that in the $J>1$ region,
and it changes substantially at $D=0$ for $J>1$ (first column).
However, the QFI becomes stable as $D$ increases (see $D=0.5, 1$).
We can also observe that the coherences increase with $\gamma$
and reach maximal values at
$\gamma=1$ for $J>1$, with the critical point occurring
at $J=1$.
The two coherence measures, namely the $l_1$ norm and
REC, behave similarly and form clear
layers for different values of $D$. 
For example, in Fig.~\ref{fig1} (ii), we can see that
as $D$ becomes larger, the
coherence becomes smaller, indicating
that higher $D$ values suppress quantum coherence.
All plots show clear differences between the two
regions $J\in[0,1]$ and $J\in[1,2]$.
The model's quantum phase transition occurs at the critical point $J=1$.
We can also see the same behavior for the
next-nearest ($r=2$) and 
fifth-nearest ($r=5$) neighbor cases.

\begin{figure*}[htbp!]
\includegraphics[width=6.3in]{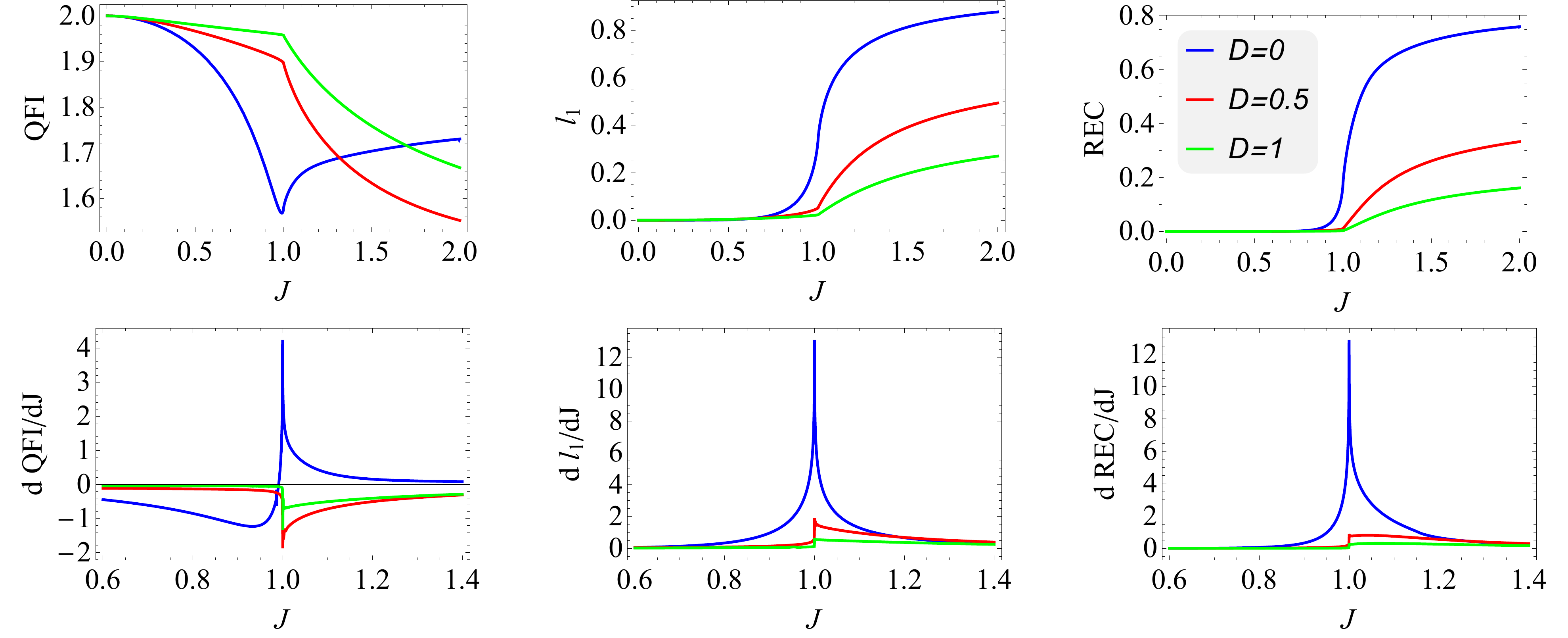}\\
\centering
\caption{(Color online) QFI, $l_1$ norm of
coherence, REC, and their derivatives vs $J$
for the fifth-nearest neighbor ($r=5$) spins in
the $\gamma=1/2$
$XY$ model with DM interactions, for $D=0,0.5,1$.}
\label{fig3}
\end{figure*}

We now investigate the $XX$ ($\gamma=0$) and
Ising ($\gamma=1$) models with
DM interactions, as shown in Figure 2 for the relationships between the
QFI, $l_1$ norm of coherence, and REC vs 
the parameters $J$ and $D$.
In each subfigure, the upper and lower surfaces represent
the nearest ($r=1$) and fifth-nearest ($r=5$) neighbor spins, respectively.
For the $XX$ model, we observe critical behavior at $J=1$.
The QFI decreases as $J$ increases in the $J>1$ region,
but the coherence increases.
The behaviors for the fifth-nearest and nearest neighbor
spin cases are identical.
For the Ising model, the QFI decreases as $J$ increases
in the nearest neighbor case. However,
it exhibits different behavior in the fifth-nearest neighbor case, where
the QFI either decreases to a minimal value or increases to a finite value.
The coherences increase with $J$.

Figure \ref{fig2} shows the QFI, $l_1$ norm of coherence,
and REC
as functions of $J$ for the nearest neighbor ($r=1$) spins in
the transverse Ising model ($\gamma=1$) with different DM interaction parameters $D$.
The first row shows the QFI, $l_1$ norm, and REC vs $J$,
whereas the second row shows the corresponding
derivative coincidences. The blue, red, and green lines correspond to DM parameters $D=0, 0.5, 1$, respectively.
Here, we notice that, similar to the QFI behavior, the coherences
have cusps at $J=1$.
In addition, they are clearly suppressed as
$D$ increases, and their derivatives clearly show
nonanalytic behavior at the critical points, implying 
that a quantum phase transition occurs at $J=1$.

Providing a contrast to Figure \ref{fig2}, Figure \ref{fig3} shows the fifth-nearest neighbor ($r=5$) spin in the $\gamma=1/2$ model. Here, when $D=0$
the QFI decreases
for $J\le1$ and increases for $J>1$.
However, the QFI exhibits different behavior
when $D=0.5$ and $1$, for which it decreases and has
a cusp at $J=1$. In addition, its
derivative is negative
when $D=0.5, 1$, and positive when
$D=0$. A quantum phase
transition can be seen clearly at $J=1$.
The coherence is initially zero, then increases
quickly near $J=1$ before saturating at
a finite value.
The derivative of the QFI and coherence exhibit nonanalytic behavior
at the critical point $J=1$, and the
DM parameter suppresses quantum phase transition.

\begin{figure*}[htbp!]
\includegraphics[width=3.6in]{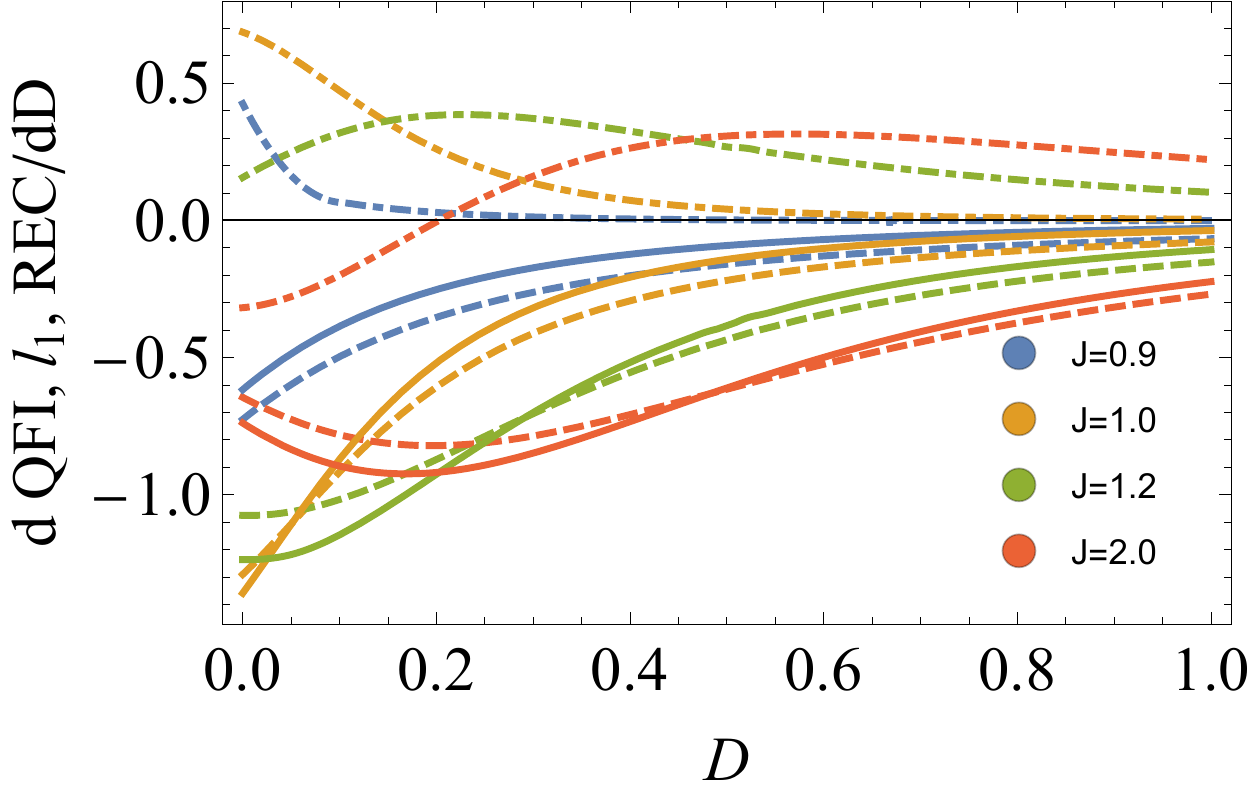}\\
\centering
\caption{(Color online) First derivatives of QFI, $l_1$ norm of coherence,
and REC with respect to $D$ for nearest neighbor spins 
($r=1$) in the $\gamma=1/2$ $XY$ model with DM
interactions. The dashed, solid, and dot-dashed lines
correspond to the QFI, $l_1$ norm, and REC
respectively, with blue, orange, green, and red lines 
denoting $J=0.9,1.0,1.2,2.0$, respectively.}
\label{fig4}
\end{figure*}

Figure \ref{fig4} shows the
derivatives of the
QFI, $l_1$ norm of coherence, and REC with respect to
the
DM interaction parameter $D$.
The dashed, solid, and dot-dashed lines correspond to
the QFI,
$l_1$ norm, and REC respectively, whereas
the blue, orange, green,
and red lines correspond to $J=0.9,1.0,
1.2,2.0$, respectively. Here,
the derivatives show distinct behaviors for
different $J$ regions. The QFI and $l_1$ norm derivatives are always negative,
but the coherence derivatives (dot-dashed lines) are
positive, except for the low $D$ region when $J=2$.
No nonanalytic behavior is observed, as was found for
quantum discord \cite{Liu2011}.

\section{Conclusions}
In this paper, we have studied quantum phase transitions
in anisotropic $XY$ chains
with DM interactions
using QFI and
quantum coherence. In particular,
we have identified quantum critical points
using QFI
and quantum coherence. We have also shown that
the coherence (characterized by
the $l_1$ norm and relative entropy)
decreases as the DM interaction increases, and that
the DM
coupling suppresses the quantum phase transition.
These results improve our understanding of
the relationships between the QFI, quantum coherence, and quantum phase
transitions in spin systems. These
quantities are important when investigating quantum phase transitions in
realistic experimental scenarios. This approach
could also be used to explore quantum
phase transitions in other spin-chain systems.

\section*{Acknowledgments}
This work is supported by the NSFC (11675113, 11765016),
NSF of Beijing under No. KZ201810028042, 
and Jiangxi Education Department Fund (GJJ161056, KJLD14088).

\end{document}